\documentclass[referee]{raa}            

\usepackage{graphicx,times}             
\usepackage{natbib}
\usepackage{amssymb,amsmath}
\bibpunct{(}{)}{;}{a}{}{,}

\usepackage[pagebackref=true]{hyperref}
\begin{document}

  \title{Producing type I\lowercase{a} supernovae from hybrid CONe white dwarfs with main-sequence binary companions at low metallicity of Z = 0.0001}

   \volnopage{Vol.0 (20xx) No.0, 000--000}      
   \setcounter{page}{1}          

   \author{Boyang Guo
      \inst{1,2,3,4}
   \and Xiangcun Meng
      \inst{1,2}
   \and Zhijia Tian
      \inst{3}
   \and Jingxiao Luo
      \inst{1,2,4}
   \and Zhengwei Liu
      \inst{1,2}
   }

   \institute{Yunnan Observatories, Chinese Academy of Sciences (CAS), Kunming 650216, China; {\it zwliu@ynao.ac.cn}\\
        \and
             International Centre of Supernovae (ICESUN), Yunnan Key Laboratory, Kunming 650216, China\\
        \and
            Department of Astronomy, Key Laboratory of Astroparticle Physics of Yunnan Province, Yunnan University, Kunming 650500, China\\
        \and
             University of Chinese Academy of Sciences, Beijing 100049, China\\
\vs\no
   {\small Received 20xx month day; accepted 20xx month day}}

\abstract{ The nature of progenitors of Type Ia supernovae (SNe~Ia) and their explosion mechanism remain unclear. It has been suggested that SNe~Ia may be resulted from thermonuclear explosions of hybrid carbon-oxygen-neon white dwarfs (CONe~WDs) when they grow in mass to approach the Chandrasekhar mass limit by accreting matter from a binary main-sequence (MS) companion. In this work, we combine the results of detailed binary evolution calculations with population synthesis models to investigate the rates and delay times of SNe~Ia in the CONe~WD~+~MS channel at low metallicity environment of $\mathrm{Z=0.0001}$. For a constant star formation rate of $5\,\mathrm{M_{\sun}\,yr^{-1}}$, our calculations predict that the SN~Ia rates in the CONe~WD~+~MS channel at low metallicity of $\mathrm{Z=0.0001}$ is about $0.11-3.89 \times 10^{-4}\,\mathrm{yr^{-1}}$. In addition, delay times in this channel cover a wide range of $0.05-2.5\,\mathrm{Gyr}$. We further compare our results to those given by previous study for the CONe~WD~+~MS channel with higher metallicity of $\mathrm{Z=0.02}$ to explore the influence of metallicity on the results. We find that these two metallicity environments give a slight difference in rates and delay times of SNe~Ia from the CONe~WD~+~MS channel, although SNe~Ia produced at low metallicity environment of $\mathrm{Z=0.0001}$ have relatively longer delay times.
}

   \authorrunning{B. Guo et al. }            
   \titlerunning{SNe I\lowercase{a} from hybrid CONe~WDs at low metallicity }  

   \maketitle

%
%
\section{Introduction}           
\label{sec:intro}

Type Ia supernovae (SNe~Ia) are the cataclysmic explosions of intermediate-mass stars at the end of their lives. SNe~Ia have typical B-band peak magnitudes of around $\mathrm{-19.5\,mag}$, which are as bright as their host galaxies. SNe~Ia have been used as the standard candles to measure cosmological parameters due to high luminosities and homogeneity of their light curves  \citep{RIES1998, PERL1999}. SNe~Ia are one of the main contributors to heavy-elements such as intermediate mass and iron-group elements in the Universe \citep[e.g.][]{MATT1986}. Despite of the importance of SNe~Ia, their progenitors and explosion mechanism remain a mystery \citep[see][for recent reviews]{MAOZ2014, LIVI2018, JHAS2019, LIUZ2023}.

It is widely accepted that SNe~Ia may be originated from thermonuclear explosions of carbon-oxygen white dwarfs (CO~WDs) in binary systems \citep[e.g.][]{HOYL1960, NOMO1982, NOMO1984, BLOO2012}. Depending on the nature of donor stars, different progenitor models have been proposed for SNe~Ia, including the single-degenerate (SD) scenario \citep[e.g.][]{WHEL1973, NOMO1984, HANZ2004}, the double-degenerate (DD) scenario \citep[e.g.][]{IBEN1984, WEBB1987}, the core-degenerate scenario \citep[e.g.][]{LIVI2003, KASH2011, ILKO2012, SOKE2013, SOKE2014, SOKE2018}, the triple model \citep[e.g.][]{KATZ2012, HAME2013, TOON2018, RAJA2023} and the single star model \citep[e.g.][]{IBEN1983, TOUT2005, ANTO2020}, etc. In this work, we will focus on the SD scenario only. In the SD scenario, the WD accretes material from its non-degenerate companion star to accumulates mass on its surface, triggering a thermonuclear explosion when the WD grows in mass to approach the Chandrasekhar-mass limit. The donor star in this scenario could be a main-sequence (MS) star, a sub-giant star, a red giant (RG) star or a helium (He) star \citep[e.g.][]{HACH1999, HANZ2004, RUIT2009, WANG2009, MENG2009, MENG2010, LIUZ2018, LIUZ2020}.

The SD scenario has advantages and disadvantages in both theoretical and observational sides, which have been comprehensively reviewed by \citet{LIUZ2023}. We therefore only briefly summarize them here. The SD scenario seems likely to explain the homogeneity of peak luminosities of the so-called normal SNe~Ia because that the WDs in this scenario are expected to explode near the Chandrasekhar-mass limit.  However, only a narrow range of mass transfer rates can lead to a stable mass accumulation onto the WD in the SD scenario. This leads to the predicted SN~Ia rates in this scenario have difficulties in explaining those inferred from observations \citep[e.g.][]{PACZ1976, FUJI1982A, FUJI1982B, LIVI1989, HANZ2004, NOMO2007, SHEN2007, RUIT2009, WANG2009,MENG2009, MENG2010, WOLF2013, CLAE2014, PIER2014, LIUZ2018, LIUZ2023}. On the observational side, there are certain observations seem to suggest that some SNe~Ia may be produced from the SD scenario. For instance, \citet{MCCU2014} reported that a blue luminous source was detected in the pre-explosion image of a SN~Iax event\footnote{SN~Iax is a subclass of SN~Ia with sub-luminosity and lower expansion velocity \citep[][]{LIWE2003, FOLE2013}.}, SN~2012Z. This blue luminous source has been thought to be a He-star companion of its progenitor system. Also, some studies reported that the narrow absorption signatures of circumstellar material that are thought to be originated from the mass transfer from the companion stars in the SD scenario have been detected in some SNe~Ia (e.g., \citealt{PATA2007, STER2011, DILD2012, SILV2013}; but see also \citealt{SOKE2013}). However, no promising surviving companion star predicted by the SD scenario has been firmly confirmed yet in nearby SN remnants (SNRs; \citealt{KERZ2009, SCHA2012, RUIZ2018}), although there are a few candidates such as Tycho~G \citep{RUIZ2004, RUIZ2019, RUIZ2023a, FUHR2005, IHAR2007, GONZ2009, KERZ2009, KERZ2013, BEDI2014}, US~708 \citep[][]{GEIE2015, BAUE2019, LIUZ2021} and MV-G272 \citep[][]{RUIZ2023b}. The features of swept-up H/He due to the SN~ejecta-companion interaction predicted by hydrodynamical models have not been firmly confirmed by the observations yet  \citep[e.g.][]{MARI2000, PAKM2008, LIUZ2012, LIUZ2013, PANK2012, BOEH2017, MCCU2022}, although there are a few candidate objects show H/He emission lines in their late-time spectra \citep[][]{LEON2007, LUND2013, LUND2015, SHAP2013, SHAP2018, MAGU2016, GRAH2015, GRAH2017, SAND2018, SAND2019, SAND2021, DIMI2017, DIMI2019, JACO2019, HOLM2019, TUCK2019, TUCK2020, TUCK2022, SIEB2020, ELIA2021, HOSS2017, HOSS2022, VALL2019, PRIE2020}. In addition, the lack of X-ray emission in current observations (see \citealt{MENG2016}, e.g. \citealt{HORE2012, MARG2014}) also seems to challenge the SD scenario.

A series of previous studies on the SD scenario generally assumed that the accreting WD is a carbon-oxygen (CO) WD  \citep[e.g.][]{HOYL1960, NOMO1984, HACH1999, HANZ2004, MENG2009, RUIT2009, WANG2014, CLAE2014, LIUZ2018}. Based on stellar evolution calculations, \citet{DENI2013} found that a CO core that goes through an off centered C-ignition in an asymptotic giant branch (AGB) star can stall its C-burning due to the convective boundary mixing. As a result, the AGB star would eject its envelope after the core nuclear reaction, leaving a hybrid CO-neon (Ne) WD consisting of a CO core and an ONe mantle \citep{DENI2013}. Furthermore, \citet{CHEN2014} suggest that the hybrid CONe~WD could be as massive as $1.3\rm\,M_{\sun}$ at a low value of the C-burning rate (CBR) factor of 0.1\footnote{The CBR factor characterizes the difficulty of C-ignition in the CO core of an AGB star \citep{DENI2013, CHEN2014}.}. Such massive CONe~WDs have been thought to likely accrete a small amount of mass from their companion stars to reach the Chandrasekhar-mass limit to trigger SN~Ia explosions if they are in binary systems \citep{DENI2015}. Therefore, some previous studies have proposed the binary systems composed of a hybrid CONe~WD and a non-degenerate companion star as a new sub-type of SD progenitor model for SNe~Ia \citep[e.g.][]{GARC1997, DENI2013, DENI2015, CHEN2014, MENG2014, WANG2014, LIUZ2015, KROM2015, MARQ2015}. Recently, \citet{KROM2015} have showed that the incomplete off-center pure deflagration explosions of near-Chandrasekhar-mass hybrid CONe~WDs (sometimes known as ``failed-detonation model''; see also \citealt[][]{JORD2012, KROM2013, FINK2014, LACH2022}) could provide reasonable agreement with the observational features of faint Type Iax SN~2008ha.
However, whether the convective boundary mixing could be strong enough to stall the c-burning inside AGB stars remains open \citep[see][]{LECO2016,LATT2017}.

By performing binary population synthesis (BPS)  calculations, \citet{MENG2014} and \citet{WANG2014} have respectively investigated the rates and delay times of SNe~Ia produced from thermonuclear explosions of near-Chandrasekhar-mass hybrid CONe~WDs with a MS donor (i.e. CONe~WD~+~MS channel) and/or a He-star donor  (i.e. CONe~WD~+~He star channel). Depending on different initial conditions and assumptions in their BPS calculations, both works have suggested SN~Ia rates from near-Chandrasekhar-mass hybrid CONe~WDs to be of the order of a few per cent of the Galactic SN~Ia rates \citep{MENG2014, WANG2014}. In addition, because that hybrid CONe~WDs are expected to form through relatively massive zero-age MS stars (ZAMS, \citealt{CHEN2014}), they found that this scenario could partly explain short delay times of SNe~Iax. However, these studies were focusing on the cases with solar-like metallicity  (i.e. $\mathrm{Z\,=\,0.02}$) only. It is unclear how different metallicities affect the SN~Ia rates and delay times in SD scenario with accreting hybrid CONe~WDs. In particular, the observations have detected SNe~Ia with high redshift of $\mathrm{z>2}$ \citep{RODN2014, RODN2015, WILL2020, HAYD2021,PIER2024}, which seem to indicate the production of SNe~Ia at lower metallicity environments. 


In this work, we investigate the SNe~Ia generated from the CONe~WD~+~MS channel at low metallicity of $\mathrm{Z=0.0001}$ by combining the results of detailed binary evolution into population synthesis calculations. We aim to study the effect of different metallicities on the SN~Ia rates and delay times in the CONe~WD~+~MS channel. The paper is organized as follows: we describe the models and methods used in this work in Section~\ref{sec:method}. The results given by our detailed binary evolution and BPS calculations are presented in Section~\ref{sec:result}. Our discussions are showed in Section~\ref{sec:discussion}. Finally, we summarize our results in Section~\ref{sec:summary}.

\section{Methods}
\label{sec:method}

\subsection{Binary evolution calculation}
\label{sec:binary} 

In this work we use the same method adopted by \citet{HANZ2004} to predict the rates and delay times of SNe~Ia produced from the hybrid CONe~WD~+~MS channel at low metallicity environment, in which the results of detailed binary evolution calculations are combined with population synthesis models. We use the stellar evolution code \texttt{Modules for Experiments in Stellar Astrophysics (MESA}, version r22.05.1; \citealt{PAXT2011, PAXT2013, PAXT2015, PAXT2018, PAXT2019}) to trace the detailed evolution of binary systems in the hybrid CONe~WD~+~MS channel. We start our binary evolution calculations when a CONe~WD~+~MS binary system is formed. In our calculations, the detailed structures of a MS companion stars is resolved consistently, and the CONe WD is treated as a point mass. The metallicity is set to be $\rm{Z\,=\,0.0001}$. This is the lowest metallicity that our codes are able to calculate. We therefore simply take $\rm{Z\,=\,0.0001}$ as an extreme case to investigate SN Ia rates and their delay times in metal-poor environments \citep[see also][]{MENG2009, CHEN2019}. We assume that the MS star transfers hydrogen-rich (H-rich) material to the WD through the Roche-lobe overflow (RLOF). If the WD grows in mass to reach the near-Chandrasekhar-mass limit (which is assumed to be $1.378\,\rm{M_{\sun}}$ in this work), we assume that it would explode as a SN~Ia \citep{NOMO1984}. We set the ratio of typical mixing length to the local pressure scaleheight, $\alpha=l/H_{p}$, to 2. In addition, the convective overshooting parameter, $\delta_{\rm{ov}}$, is set to 0.12 \citep{POLS1997, SCHR1997}, which means that the overshooting length is approximately equal to 0.25 pressure scaleheights ($H_{p}$).

In this work, we focus on the hybrid CONe~WD+MS channel for SNe Ia. In \citet{CHEN2014}, they suggest that a hybrid CONe WD forms when a massive AGB star stalls its CO core burning due to the convective boundary mixing. They further showed that the formed CONe WDs have a range of masses of 1.02\,--\,1.30 $\rm{M_{\sun}}$ for a range of CBR factors of 0.1\,-\,10. Following the results of \citet{CHEN2014}, we therefore set a mass range of our initial CONe~WDs to 1.0\,-\,1.3 ${\rm{M_{\sun}}}$ in our detailed binary evolution calculations. It is important to note that the CBR during the C burning process is quite uncertain. For instance, taking the possible resonance and hindrance effects \citep{JIAN2007, SPIL2007} into account could cause the values of the CBR change by a few orders of magnitude \citep{CHEN2014}. Therefore, we treat the CBR as a free parameter in our calculations and set it to be the same values as used in \citet{CHEN2014}.

As the MS companion star evolves, it expands to fill its Roche lobe during either the MS phase or the Hertzsprung gap (HG) phase. As a result, the companion star starts to transfer H-rich material to the WD through RLOF. As described in \citet{HANZ2004}, we also adopt the prescription of \citet{HACH1999} for the accumulation of transferred H-rich material onto the CONe~WD (i.e. the mass growth of a CONe~WD). If the mass-transfer rate ($|\dot{M}_{2}|$) is above the critical rate ($\dot{M}_{\rm{cr}}$), we assume that the transferred H burns steadily on the surface of WD and that the accreted H-rich material converts into He at a rate of $\dot{M}_{\rm{cr}}$, and the unprocessed H-rich material will lose from the binary system in optically thick wind. Here, the critical mass-transfer rate is given as follows \citep[see also][]{HACH1999}.
\begin{equation}
\dot{M}_{\rm{cr}}\,=\,5.3\,\times\,10^{-7}\dfrac{(1.7\,-\,X)}{X}(M_{\mathrm{WD}}\,-\,0.4),
\end{equation}
where $M_{\rm{WD}}$ is the mass of the accreting WD (mass is in $\mathrm{M_{\sun}}$) and $X$ is the H mass fraction.

If $|\dot{M}_{2}|$ is lower than $\dot{M}_{\rm{cr}}$, different assumptions are adopted for the mass growth of the accreting WD: (i) when $|\dot{M}_{2}|$ is higher than $\frac{1}{8}\dot{M}_{\rm{cr}}$, we assume that the transferred H-rich material will burn steadily on the WD surface or that the weak H flashes occur, and all the transferred material would retain onto the WD. (ii) When the $|\dot{M}_{2}|$ is lower than $\frac{1}{8}\dot{M}_{\rm{cr}}$, H-shell flashes are too strong to accumulate any transferred material. Therefore, the accumulation rate of the He-shell of the accreting WD can be described as follows.
\begin{equation}
\dot{M}_{\mathrm{He}}\,=\,\eta_{\mathrm{H}}|\dot{M}_{2}|,
\end{equation}
where
\begin{equation}
	\eta_{\mathrm{H}} = \left\{
	\begin{aligned}
	&\dot{M}_{\rm{cr}}/|\dot{M}_{2}|, & ~ & |\dot{M}_{2}|>\dot{M}_{\rm{cr}},\\
	&1,  & ~ & \dot{M}_{\rm{cr}}\geq|\dot{M}_{2}|\geq\frac{1}{8}\dot{M}_{\rm{cr}},\\
	&0,  & ~ & |\dot{M}_{2}|<\frac{1}{8}\dot{M}_{\rm{cr}}.
	\end{aligned} 
	\right.
\end{equation}

We further assume that the He ignites when the He layer reaches a certain mass and that He is converted into CO. The He flashes may occur during the He-shell burning and a portion of the shell mass is assumed to be blown away in our calculations. In this work, the mass accumulation efficiency for He-shell flashes is taken from \citet{HACH1999}, which can be described as follows.

\begin{equation}
	\eta_{\mathrm{He}} = \left\{
	\begin{array}{lc}
	-0.175(\mathrm{log_{10}}\,\dot{M}_{\mathrm{He}}+5.35)^{2}+1.05,\\
	\hspace{3.0cm} -7.3<\mathrm{log_{10}}\,\dot{M}_{\mathrm{He}}<-5.9,\\
	1,  \hspace{2.7cm} -5.9\le \mathrm{log_{10}}\,\dot{M}_{\mathrm{He}}\le-5,
	\end{array} 
	\right.
\end{equation}

Finally, we can calculate the mass growth rate (i.e. $\dot{M}_{\mathrm{WD}}$) of an accreting CONe~WD, which is  
\begin{equation}
\dot{M}_{\mathrm{WD}}\,=\,\eta_{\mathrm{He}}\dot{M}_{\mathrm{CO}}\,=\,\eta_{\mathrm{He}}\eta_{\mathrm{H}}|\dot{M}_{2}|.
\end{equation}

The mass lost from the binary system is assumed to take away the specific orbital angular momentum of the accreting WD. According to the results of \citet[][see their Figure~5]{CHEN2014}, we consider the cases with different CBR factors of 0.1, 1.0 and 10.0, and set the initial mass of CONe~WD ($M_{\mathrm{WD}}^{i}$) ranges from 1.0 to $1.3\,\rm{M_{\sun}}$. For the WDs with a mass range of $1.0\,$--$\,1.2\,\mathrm{M_{\sun}}$, we set the initial mass range of the donor star ($M_{2}^{i}$) to $1.0\,-\,3.6\,\mathrm{M_{\sun}}$, which is consistent with that in \citet{MENG2014}. For the cases with a WD mass of 1.3$\,\mathrm{M_{\sun}}$, we range the donor mass $M_{2}^{i}$ from $1.0\,\,\mathrm{M_{\sun}}$ to $\,4.0\,\mathrm{M_{\sun}}$. In addition, we set a range of initial orbital periods ($P_{i}$) to $0.25\,$--$\,20\,$days.

\subsection{Binary population synthesis}
\label{sec:bps}

In order to calculate the birth rates and delay times of SNe~Ia produced from the CONe~WD~+~MS channel, we perform a series of population synthesis calculations with the rapid binary evolution code developed by \citet{HURL2000, HURL2002}, i.e., the \texttt{Binary Star Evolution (BSE)} code. The \texttt{BSE} code can evolve a large number of binary systems rapidly from ZAMS to the formations of CONe~WD~+~MS binary systems. The basic assumptions and treatments for some fundamental processes in our population synthesis calculations are similar to those in \citet{MENG2014}, which are briefly described as follows.

\subsubsection{Formation of CONe~WD~+~MS systems}

In Figure~5 of \citet[][]{CHEN2014}, they have provided mass boundaries that allow to form CO~WDs, ONe~WDs and hybrid CONe~WDs for different CBR factors. Usually, CO cores in the AGB stars with significantly large masses will go through a C burning process and evolve into a ONe core. However, according to \citet{DENI2013}, the C burning process may stall due to the convective boundary mixing and form a hybrid CONe core. Since the convective boundary mixing process is not considered in the \texttt{BSE} code, hybrid CONe WDs will be mis-evaluated into ONe~WDs in our rapid binary evolution calculations. We therefore simply assume that a WD would be a hybrid CONe~WD if its mass is lower than the upper-limit mass boundaries of hybrid CONe~WDs given by Figure~5 of  \citet[][the upper-panel]{CHEN2014} and appear to be a ONe~WD in the binary population synthesis \citep[see also][]{MENG2014}.

Based on the results of \citet{CHEN2014}, we performed BPS calculations for three different CBR factors of 0.1, 1 and 10. If a CONe~WD~+~MS binary system is formed and its binary parameters at the onset of RLOF fall into the initial contours in $\mathrm{log_{10}}\,P^{i}\,$--$\,M_{2}^{i}$ plane given by our detailed binary evolution calculation in Section~\ref{sec:binary}, we assume that this system would eventually explode as a SN~Ia.

\subsubsection{Treatment for common-envelope evolution}

Different binary systems start their first RLOF in different stages, which depends on the mass, mass ratios and orbital periods. Whether or not a binary system can lead to dynamically unstable mass transfer to form a CE mainly depends on the mass ratio of the system at the onset of RLOF. If the mass ratio is larger than the critical mass ratio, $q_{c}$, the dynamically unstable RLOF will happen to form a CE. Here, the critical ratio $q_{c}$ varies with the evolutionary stages of the binary systems at the onset of RLOF \citep{HJEL1987, WEBB1988, HANZ2002, PODS2002, CHXF2008}. Based on the results of previous detailed binary evolution studies \citep[e.g.][]{HANZ2000, CHXF2002, CHXF2003}, we set $q_{c}=4.0$ when the primary star is either on the MS phase or on the HG phase. If the primary star is a naked He giant, we set $q_{c}\,=\,$0.748. If the primary star is on the first giant branch or AGB, we set $q_{c}$ as follows.
\begin{equation}
q_{c}~=~\left[1.67\,-\,x\,+\,2\left(\frac{M_{\rm{c1}}^{\rm{P}}}{M_{1}^{\rm{P}}}\right)\right]\,/\,2.13
\end{equation} The filled circles and cross symbols respectively indicate that the binary
systems that can and cannot successfully lead to SNe Ia. Where $M_{\,1}^{P}$ and $M_{\rm{c1}}^{P}$ are the mass and the core mass of the primary star, respectively; $x=\mathrm{d\,ln}\,R_{1}^{P}/\mathrm{d\,ln}\,M_{1}^{P}$ is the mass-radius exponent of the primary star, which varies with composition.

Binary systems inside the CE are expected to release their orbital energies through the frictional drag and then undergo an in-spiral. The released energy would be partly used to eject the CE. Once a binary system enters the CE phase, we assume that the CE could be ejected completely if
\begin{equation}
\alpha_{\rm{CE}}~\Delta E_{\rm{orb}}~=~|E_{\rm{bind}}|,
\end{equation}
where $\Delta E_{\rm{orb}}$ is the orbital energy released during the spiral-in phase, $E_{\rm{bind}}$ is the binding energy of the CE, and $\alpha_{\rm{CE}}$ is the CE ejection efficiency. Because that the internal energy in the CE is not incorporated into the binding energy, $\alpha_{\rm{CE}}$ may be greater than 1. The exact value of $\alpha_{\rm{CE}}$ is still poorly constrained from both theoretical and observational sides \citep[e.g.][]{ZORO2010,TOON2013,IVAN2013,SCHE2023,ROPK2023,GEHO2024}. In this work, similar to various previous works \citep[e.g.][]{MENG2009, MENG2014, MENG2018,LIUZ2018,WANG2013,WANG2014,WANG2017}, we simply set $\alpha_{\rm{CE}}$ to be 0.5, 0.75, 1.0 and 3.0, to investigate the influence of different CE ejection efficiencies on the results. $E_{\rm{bind}}$ is given by
\begin{equation}
E_{\rm{bind}}~=~\frac{\rm{G}M_{\rm{1}}^{\rm{P}}M_{\rm{e1}}^{P}}{\lambda_{\rm{CE}}\,R_{1}^{\rm{P}}},
\end{equation}
Where G is the gravitational constant, and $M_{c1}^{\rm{P}}$ is the mass of the removed envelope from the giant primary. Here $M_{\rm{e1}}^{\rm{P}}\,=\,M_{1}^{\rm{P}}-M_{\rm{\rm{c1}}}^{\rm{P}}$. $R_{1}^{\rm{P}}$ is the radius of the primary giant star at the onset of CE, while $\lambda_{\rm{CE}}$ is the binding energy factor, which is set to characterize the envelope central concentration of the giant primary. In this work, the binding energy factor is set to be $\lambda_{\rm{CE}}=0.5$.

\subsubsection{Basic assumptions}

We use a Monte Carlo method to generate a large number of binary systems and follow their rapid evolution from ZAMS phase to the formation of CONe~WD~+~MS systems. The basic assumptions adopted in our rapid calculations are described as follows:

(1) We assume that all stars are in binary systems. Also, all binary systems are assumed to have a circular orbit. 

(2) We use the initial mass function (IMF) given by \citet{MILL1979}. The primary star is generated according to the formula of \citet{EGGL1989}:
\begin{equation}
M_{1}^{P}~=~\frac{0.19~R}{(1-R)^{0.75}~+~0.0032~(1~-~R)^{0.25}},
\end{equation}
where $R$ is a random number uniformly distributed in the range [0,1], $M_{1}^{P}$ is the mass of the primary star, which is between $0.1\,\mathrm{M_{\sun}}$ and $100\,\mathrm{M_{\sun}}$.

(3) We assume that the mass ratios of the initial binary systems at ZAMS phase, $q'$, have a constant distribution \citep{MAZE1992, GOLD1994}.
\begin{equation}
n(q')~=~1, \hspace{2.cm} 0~<~q'\leq1,
\end{equation}
where $q'~=~M_{1}^{P}~/~M_{2}^{P}$.

(4) We assume that the distribution of binary separations is constant in $\mathrm{log}\,a$ for wide binary systems(where a is separation), and falls off smoothly for close binary systems \citep{HANZ1995}:
\begin{equation}
a\cdot n(a)=\left\{
\begin{array}{lc}
\alpha_{\rm{sep}}(a/a_{\rm{0}})^{\rm{m}} & a\leq a_{\rm{0}},\\
\alpha_{\rm{sep}}, & a_{\rm{0}}<a<a_{\rm{1}},\\
\end{array}\right.
\end{equation}
where $\alpha_{\rm{sep}}\,\approx\,0.07$, $a_{0}\,=\,10\, \rm{R_{\sun}}$, $a_{1}\,=\,5.75~\times~10^{6}\,\rm{R_{\sun}}\,=\,0.13\,\mathrm{pc}$ and $m\,\approx\,1.2$. This distribution implies that the number of wide binary systems per logarithmic interval is equal and about 50 per cent of binary systems have orbital periods less than $\rm{100\,yr}$ \citep{HANZ1995}. 

(5) Either a single starburst with $10^{11}\,\mathrm{M_{\sun}}$ or a constant star formation rate (SFR) of $5\,\mathrm{M_{\sun}\,yr^{-1}}$ in the past $\rm{15\,Gyr}$ is assumed in our calculations.

\section{Results}
\label{sec:result}

In this section, we present the main results from our detailed binary evolution and population synthesis calculations, including the initial parameter spaces (i.e. the orbital period-secondary mass plane; $\mathrm{log_{10}}\,P^{i}-M_{2}^{i}$) which eventually lead to SNe~Ia in the hybrid CONe~WD~+~MS channel, and the predicted birth rates and delay times of SNe~Ia from this channel.

\subsection{Initial parameter contours}
\label{sec:contours} 

\begin{figure*}
    \centering
	\includegraphics[width=0.9\linewidth]{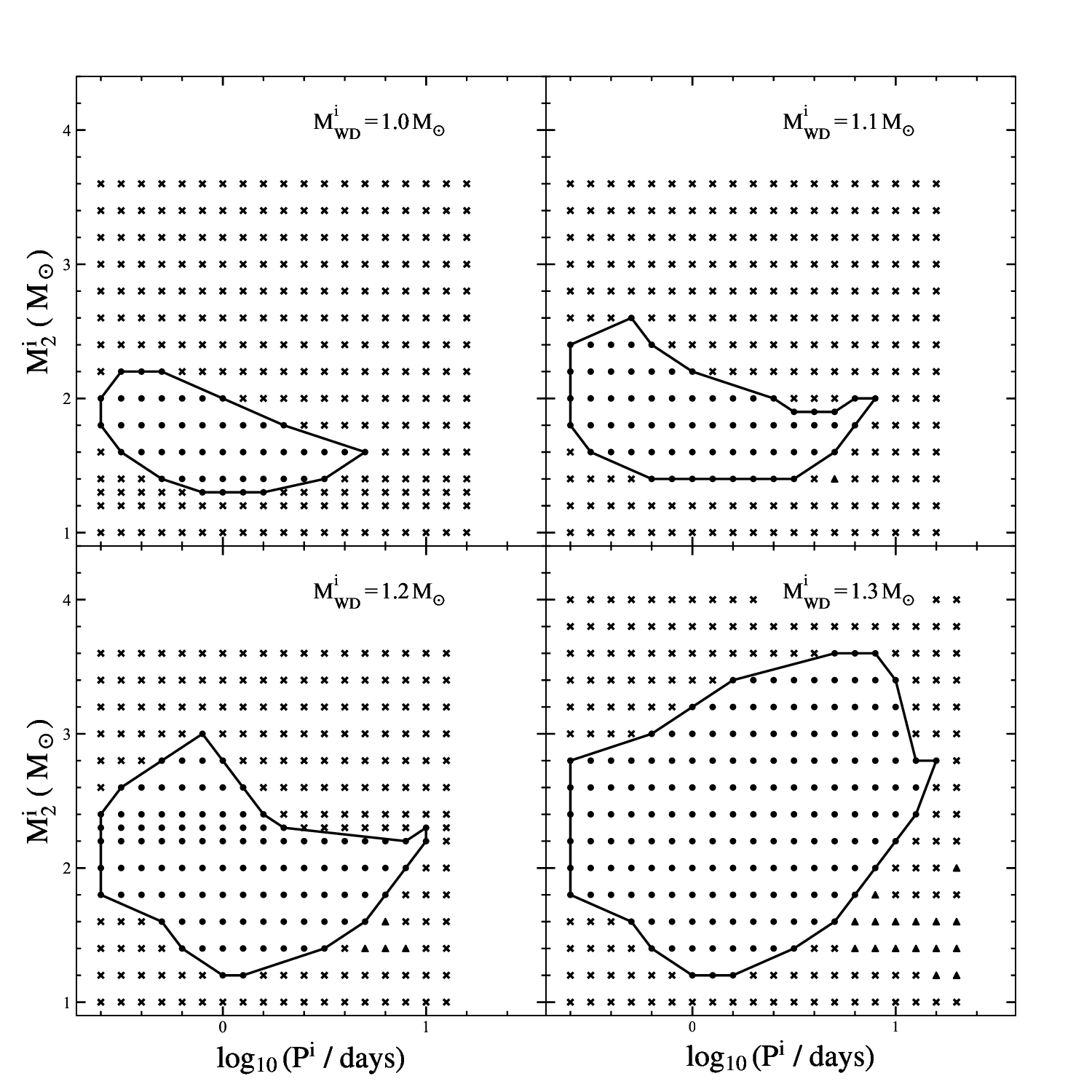}
	\captionsetup{font={footnotesize}}  
	\caption{Results in the orbital period–secondary mass ($\mathrm{log_{10}}P^{i}-M_{2}^{i}$) plane for CONe~WD~+~MS channel at $\mathrm{Z=0.0001}$ from our detailed binary evolution calculations. Here, $P^{i}$ and $M_{2}^{i}$ are the initial orbital period and companion mass, respectively. The filled circles and cross symbols respectively indicate the binary systems that can and cannot successfully lead to SNe~Ia. Different panels correspond to the results with different initial CONe~WD masses ($M_{\rm WD}^{i}$).}
	\label{fig:1}
\end{figure*}

We use the \texttt{MESA} code to perform detailed binary evolution calculations for about 1000 binary systems consisting of a CONe~WD (which is treated as a point mass) and a MS companion star by using the method described in Section~\ref{sec:binary}. The initial masses of CONe~WDs and MS companions and orbital periods of these binary systems cover the ranges of $1.0\,\mathrm{M_{\sun}}\leq M_{\mathrm{WD}}^{i} \leq 1.3\,\mathrm{M_{\sun}}$, $1.0\,\mathrm{M_{\sun}}\leq M_{\mathrm{2}}^{i} \leq 4.0\,\mathrm{M_{\sun}}$ and $-0.6\leq \mathrm{log_{10}}\,P^{i} \leq1.3$, respectively. Figure~\ref{fig:1} presents the results from our detailed binary evolution calculations in the $\mathrm{log_{10}}\,P^{i}-M_{2}^{i}$ plane for different given initial WD masses. The filled circles and cross symbols in Figure~\ref{fig:1} respectively correspond to the models that are able and unable to successfully produce SN~Ia explosions in our calculations. \textit{The left boundaries} of initial contours (solid lines) in Figure~\ref{fig:1} are constrained by the requirement of that the binary systems are not filling their Roche lobes before the MS phase. Binary systems outside \textit{the upper boundaries} of the contours subsequently experience dynamically unstable mass transfer and lead to CE objects. For binary systems outside \textit{the right boundaries}, their mass transfer rates are too high to lead to a long enough mass transfer phase. This inhibit the WDs to accumulate mass to reach the Chandrasekhar limit\footnote{Note that a small number of systems with a high initial WD mass ($M^{i}_{\rm WD}\rm{\geq1.1\,M_{\sun}}$) and a low initial MS mass ($M^{i}_{2}\rm{\leq2.0\,M_{\sun}}$) outside the right boundaries can still undergo dynamically stable mass transfer when the donor star evolve to the base of red giant branch (RGB) to eventually lead to SNe~Ia. In this work, however, we only focus the hybrid CONe~WD~+~MS channel.}. For binary systems below \textit{the lower boundaries} of the contours, the MS donors are not massive enough to lead to  sufficient mass-transfer rates to allow the WDs to accumulate mass to reach the Chandrasekhar limit.

\begin{figure*}
    \centering
	\includegraphics[width=0.9\linewidth]{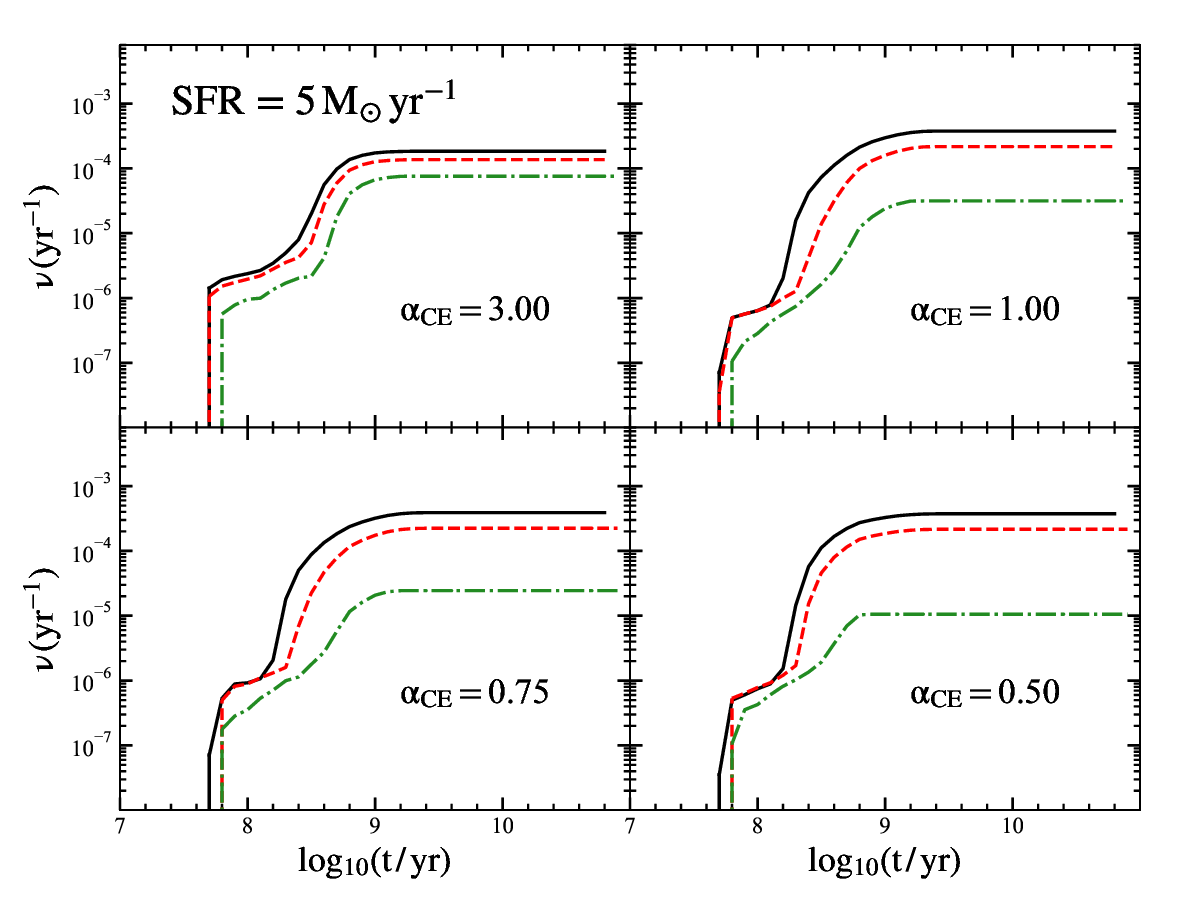}
	\captionsetup{font={footnotesize}} 
	\caption{Evolution of SN~Ia birth rates from the CONe~WD~+~MS channel for a constant star formation rate of $\mathrm{SFR}=5\,\mathrm{M_{\sun}\,yr^{-1}}$. The results with different CE ejection efficiencies ($\alpha_{\rm{CE}}=0.50, 0.75, 1.00$ and 3.00) are indicated in each panel. The solid, dashed and dash-dotted lines represent the cases for CBR factors of 0.1, 1, and 10 \citep[][]{CHEN2014}, respectively.}
	\label{fig:2}
\end{figure*}

\begin{figure*}
    \centering
	\includegraphics[width=0.9\linewidth]{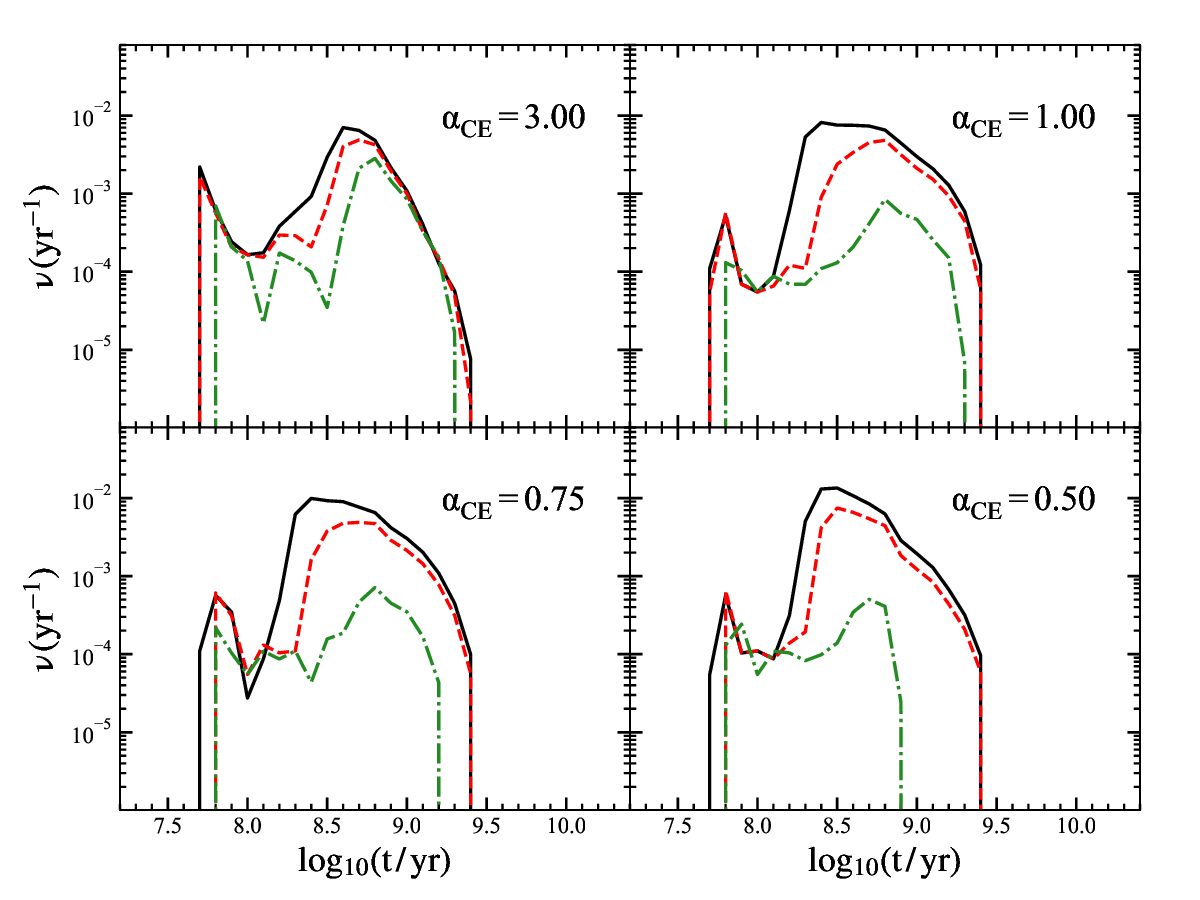}
	\captionsetup{font={footnotesize}} 
	\caption{Similar to Figure~\ref{fig:2}, but for a single starburst of $10^{11}\,{M_{\sun}}$.}
	\label{fig:3}
\end{figure*}

As shown in Figure~\ref{fig:1}, the contours for the initial CONe~WD mass of $1.3\,\mathrm{M_{\sun}}$ extends to a significantly higher initial donor mass (the upper boundary) comparing to those of other models, i.e., the enclosed region of this contour is much larger. This is because that binary systems close to the upper boundaries may go through a short dynamically stable mass transfer phase before the dynamically unstable mass transfer starts. This phase will only allow a small amount of mass accumulation ($\sim0.1\,\mathrm{M_{\sun}}$) onto the CONe~WD due to the short time scale. Thus, only a $1.3\,\mathrm{M_{\sun}}$ CONe~WD can accumulate enough mass to reach a mass of $1.378\,\mathrm{M_{\sun}}$ to trigger an SN~Ia explosion.

\subsection{Results of binary population synthesis}
\label{sec:BPS result}

By using the method described in Section~\ref{sec:bps}, we follow a rapid evolution of $10^{8}$ binary systems from the ZAMS phase to the formation of CONe~WD~+~MS systems. Once the formed CONe~WD~+~MS systems fall into the initial contours given by detailed binary evolution calculations(i.e. Figure~\ref{fig:1}) at the moment of RLOF, we simply assume that the systems would eventually lead to successful SN~Ia explosions. In Figure~\ref{fig:2}, we present the predicted birth rates of SNe~Ia from our BPS calculations for the hybrid CONe~WD~+~MS channel by assuming a constant star formation rate (SFR) of $\sim5\,\mathrm{M_{\sun}}\,\rm{yr^{-1}}$ over the past 15~Gyr. As shown in Fig.~\ref{fig:2}, the SN~Ia rate in a galaxy with a constant SFR is affected not only by the assumed CE ejection efficiencies in our BPS calculations but also by the adopted CBR factors. Depending on different $\mathrm{\alpha_{CE}}$ and CBR factors, we find that the SN~Ia rate in a galaxy with a constant SFR in the hybrid CONe~WD~+~MS channel at $\rm Z=0.0001$ is about $0.11\,$--$\,3.89 \times \mathrm{10^{-4}\,yr^{-1}}$. In addition,  Fig.~\ref{fig:2} shows that a lower $\alpha_{\rm{CE}}$ tends to give a higher SN~Ia rate. This is because that the lower the $\alpha_{\rm{CE}}$ values are, the more orbital energy will be lost when the primordial binary systems ejecting the CE to form a CONe~WD~+~MS system. As a result, the CONe~WD~+~MS systems will have shorter initial orbital separations. This refers to shorter initial Roche lobe radius, thus more easily to start the RLOF,  leading to higher SN~Ia rates.

Figure~\ref{fig:3} presents the delay-time distributions (DTDs) of SNe~Ia in the hybrid CONe~WD~+~MS channel at $\rm Z=0.0001$ after a starburst of  $10^{11}\mathrm{M_{\sun}}$. We find that the delay times of SNe~Ia in this channel cover a range of $0.05\,$--$\,2.5\,\mathrm{Gyr}$. Two peaks are observed in DTDs of SNe~Ia, which is in consistent with the result shown in \citet{MENG2014}. The right peak, around $0.3\,\mathrm{Gyr}$, is given by SNe~Ia produced from normal CONe~WD~+~MS evolutionary channel. The left peak, around $0.06\,\mathrm{Gyr}$, is from the so-called He-enriched MS (HEMS) evolutionary channel \citep[see][for a detailed description]{MENG2009, LIUZ2018}. In the HEMS channel, the primary star begins its first episode of RLOF mass transfer during the HG phase or during RG phase. Binary system then undergoes a CE phase due to the dynamically unstable mass transfer. After the CE ejection, the primary He star and a companion MS star remained in the system. As the primary He star continues to evolve and expand, the second episode of RLOF mass transfer occurs. In this phase, He-rich material transfers on to the surface of companion MS star and eventually form a CONe~WD~+~HEMS binary system \citep{LIUZ2018}. SNe~Ia formed in the hybrid CONe~WD~+~HEMS channel have relatively shorter delay times compared to those of normal hybrid CONe~WD~+~MS channel.

\begin{figure*}
    \centering
	\includegraphics[width=0.9\linewidth]{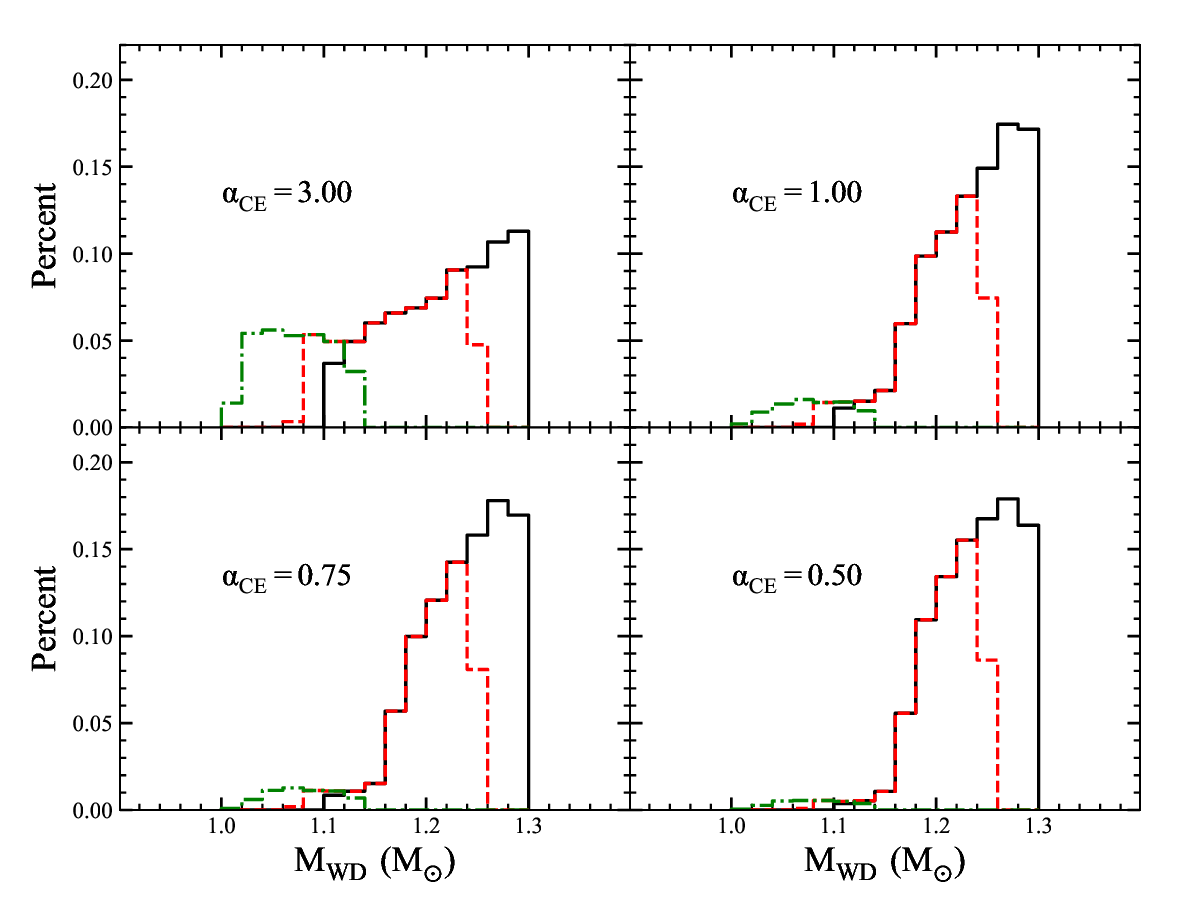}
	\captionsetup{font={footnotesize}} 
	\caption{Distributions of the initial masses of hybrid CONe~WDs from our population synthesis calculations with different CBR factors and CE ejection efficiencies ($\alpha_{\rm{CE}}$). Different lines indicate the same CBR factors as in Figs.~\ref{fig:2} and~\ref{fig:3}.}
	\label{fig:4}
\end{figure*}

Figure~\ref{fig:4} shows the mass distribution of the initial hybrid CONe~WDs of the systems which successfully produce SNe~Ia. The lower- and upper-limits for the initial CONe~WD masses (which are about $1.02\,\mathrm{M_{\sun}}$ and $1.30\,\mathrm{M_{\sun}}$) are respectively constrained by the upper limits for CO~WD and CONe~WD masses given by Figure~5 of \citet{CHEN2014}. The difference of distribution between CBR factor of 0.1 and 1 is relatively small, leading to that the SN~Ia rates of these two cases are quite similar  (see Figs.~\ref{fig:2} and~\ref{fig:3}). For a CBR factor of 10, however, more CONe~WD~+~MS systems that fulfill the condition for SN~Ia production are formed as the $\alpha_{\rm{CE}}$ factors increase, as well as the SN~Ia rates shown in Fig.~\ref{fig:2}.

\begin{figure}
    \centering
	\includegraphics[width=0.9\linewidth]{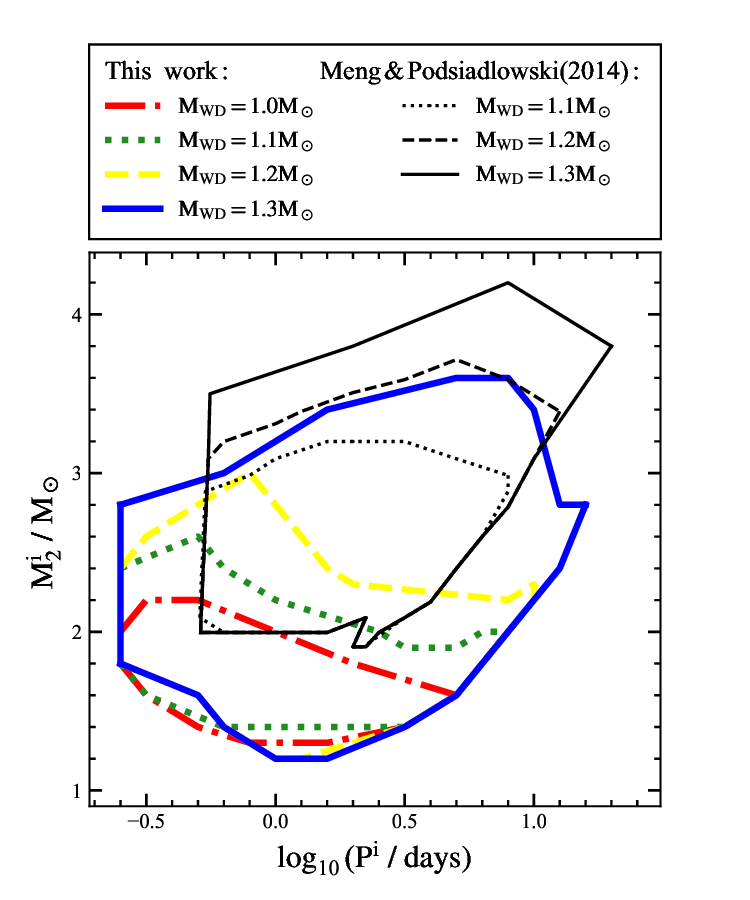}
	\captionsetup{font={footnotesize}} 
	\caption{Comparison of initial parameter spaces between our work (\textit{thick color lines}) and previous study of \citet{MENG2014} with $\mathrm{Z=0.02}$ (\textit{thin black lines}) . Results with different initial CONe~WD masses are indicated by different line-styles.}
	\label{fig:5}
\end{figure}

\section{Discussion}
\label{sec:discussion}

\subsection{Comparisons with previous works}

\begin{figure*}
    \centering
	\includegraphics[width=0.9\linewidth]{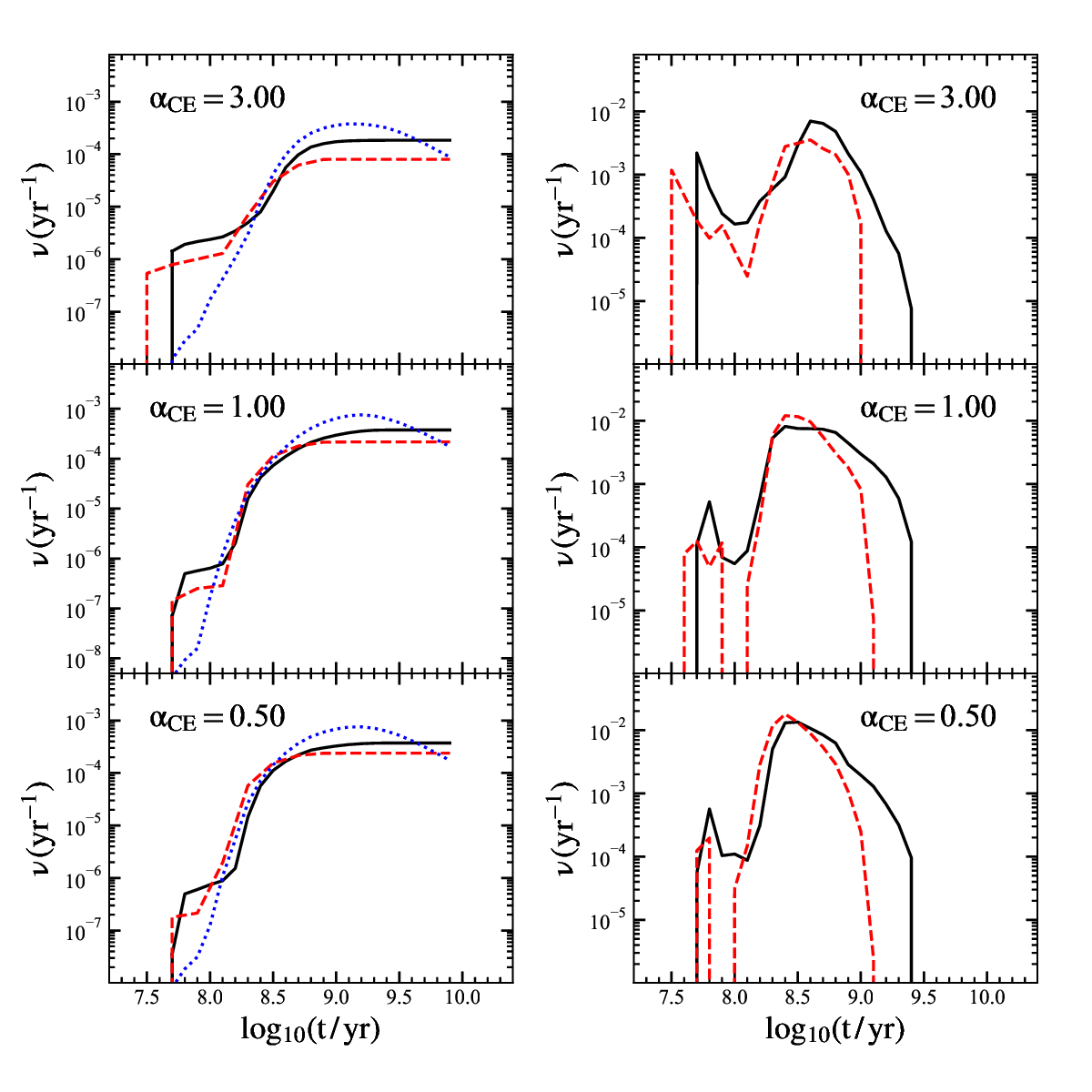}
	\captionsetup{font={footnotesize}} 
	\caption{Comparison of birth rates and delay times of SNe~Ia in a galaxy with a constant SFR(left panels) and in a starburst galaxy (right panels). Black solid lines are the results from} this work with a constant SFR model, blue dotted lines are the results from this work with a variable SFR model adopted from \citet{KUBR2015} and red dashed lines are the results from previous work of \citet{MENG2014} with $\mathrm{Z=0.02}$. Here, only the results with a CBR factor of 0.1 are given for a comparison.
	\label{fig:6}
\end{figure*}

In Figure~\ref{fig:5}, we compare the initial contours in the $\mathrm{log_{10}}\,P^{i}-M_{2}^{i}$ plane from our detailed binary evolution calculations ($\mathrm{Z=0.0001}$) to those given by previous work of \citet{MENG2014} with $\mathrm{Z=0.02}$. It shows that our contours cover shorter orbital periods and lower MS companion masses than theirs. Similar trends are also shown in the binary systems from the CO~WD~+~MS channel in \citet{MENG2009}. In Figure~\ref{fig:6}, we further compare the predicted SNe~Ia rates in a galaxy with a constant SFR and delay times of SNe~Ia between two metallicities. Here, only the results with a CBR factor of 0.1 are presented. It shows that the SNe~Ia rates in a galaxy with a constant SFR does not change significantly as the metallicity varies from $\mathrm{Z=0.02}$ to $\mathrm{Z=0.0001}$. However, Figure~\ref{fig:6} shows that SNe Ia have relatively longer delay times in lower-metallicity environments. This is because that progenitor systems in lower-metallicity environments have smaller initial secondary masses (see Fig.~\ref{fig:5}), leading to longer evolutionary timescales until they fill their Roche-lobe and thus longer delay times \citep[see also][]{MENG2009}.

To compare our results with the observations, following the previous work of \citet{KIST2013}, we simply calculate the galactic masses in a low-metallicity galaxy with Z=0.0001 by using the relation between stellar metallicities and galactic masses derived from Sloan Digital Sky Survey (SDSS) observations \citep{GALL2005}. This gives a galactic mass of $\sim2.5\,\times\,10^{7}\,\rm{M_{\sun}}$ for Z=0.0001. We then the observational relation between SNe Ia birth rates and stellar masses provided by \citet{WISE2021} to calculate the expected SN Ia rates at Z=0.0001, which is $\sim6.6\,\times\,10^{-5}\,\rm{yr^{-1}}$. These rates are relatively lower than those from our BPS calculations of $\sim\rm 3.9\,\times\,10^{-4}\,yr^{-1}$ by a factor of 6. However, it is important to note that BPS calculations have some uncertainties \citep[e.g.][see also Sect.~\ref{sec:uncertainties}]{Han2020RAA}. Also, we directly use the relations given by \citep{GALL2005} and \citet{WISE2021} for calculations under some assumptions such as that the effect of redshift could be neglected. This may lead to some uncertainties on the expected SN Ia rates at Z=0.0001.

\subsection{Uncertainties of our results}
\label{sec:uncertainties}
In this work, we simply assume that all stars are in binary systems. This means our calculations only present upper limits of the SNe Ia birth rates. We also assume that all the binary systems are in circular orbits. This assumption may lead to some uncertainties on the results. In previous work of \citet{WANG2013}, they investigated the effect of different initial eccentricities on their BPS results. They suggested that different initial eccentricities lead to a small difference in rates and delay times of SNe Ia from their BPS calculations comparing with those with an assumption of a circular orbit. In addition, we simply use a constant SFR of $\sim5\,\rm{M_{\sun}}$ for our calculations, which could lead to that SN Ia rates in Figure \ref{fig:6} are under-or over-estimated. For a comparison, we adopt a variable SFR given by \citet{KUBR2015} to study its effect on SN Ia rates. As shown in Figure \ref{fig:6}, with a variable SFR, SN Ia rates rise and reach a peak at delay times around 1\,Gyr, then decline exponentially. This indicates that different SFRs are needed to take into account in BPS studies for SNe Ia.

We adopt the optically thick wind model \citep{HACH1996, HACH1999, KATO1999} for the treatment of the mass accumulation efficiency onto the WDs.  However, the exact mass accumulation efficiency is still not well-constrained, which might bring some uncertainties on our results. Previous BPS studies have shown that adopting different mass accumulation efficiencies onto the WDs in BPS calculations could give a significant difference on the predicted SN~Ia rates \citep[e.g.][]{PIER2014, RUIT2014, TOON2014}.

We have tested the influence of different CE ejection efficiencies on the results by varying the values of $\alpha_{\rm{CE}}$ from 0.5 to 3.0 in our BPS calculations. However, current constraints on the CE ejection efficiency are still quite weak. Meanwhile, a fixed binding energy factor value ($\lambda_{\rm{CE}}\,=\,0.5$) is used for all evolutionary phased in a single BPS calculations \citep[see also][]{MENG2014}. \citet{IVAN2011} suggested that the value of $\lambda_{\rm{CE}}$ is dependent to the relative mass distribution of the envelope of the donor stars, and thus leading to that the exact $\lambda_{\rm{CE}}$ value might vary with different evolutionary phases. In addition, other initial condition and assumptions in BPS calculations such as the SFR and initial mass function are still quite uncertain \citep[see][for a detailed description of theoretical uncertainties in BPS studies of SNe~Ia]{CLAE2014}. This may also lead to some uncertainties of our BPS results.

\section{Summary}
\label{sec:summary}

In this work, we have addressed the rates and delay times of SNe~Ia produced from the hybrid CONe~WD~+~MS channel at low metallicity of $\rm Z=0.0001$ by combining the results of detailed binary evolution calculation with \texttt{MESA} \citep[][]{PAXT2011, PAXT2013, PAXT2015, PAXT2018, PAXT2019} into the population synthesis models computed from \texttt{BSE} \citep{HURL2000, HURL2002}. We adopt the `optically thick wind model' for the treatment of the mass accumulation efficiency onto the accreting CONe~WD. Comparing with previous works for solar-like metallicity of $\rm Z=0.02$ \citep{MENG2014}, we attempt to explore how different metallicities affect the rates and delay times of SNe~Ia in the hybrid CONe~WD~+~MS channel. Our main results and conclusions can be summarized as follows. 

\begin{enumerate}

\item[(1)] Depending on different CE ejection efficiencies and CBR factors adopted in BPS calculations, we find that the SN~Ia rates in a galaxy with a constant SFR of $\sim5\,\rm{M_{\sun}}$ from the hybrid CONe~WD~+~MS channel at low metallicity of $\rm Z=0.0001$ are  $(0.11-3.89)\times\mathrm{10^{-4}\,yr^{-1}}$.

\item[(2)] The DTDs of SNe~Ia from the hybrid CONe~WD~+~MS channel at $\rm Z=0.0001$ cover a wide range of $0.05\,$--$\,2.5\,\mathrm{Gyr}$, and most SNe~Ia have delay times of $0.3\,\mathrm{Gyr}$.

\item[(3)] Comparing with previous works of \citet{MENG2014} for $\rm Z=0.02$, we find that SN~Ia rates and delay times given by our work are not significantly different from theirs, although SNe~Ia produced at low metallicity environment of $\mathrm{Z=0.0001}$ have relatively longer delay times. 

\end{enumerate}

\begin{acknowledgements}
We are grateful to Hai-Liang Chen, Zhenwei Li, Yunlang Guo and Lifu Zhang for their fruitful discussions. This work is supported by the National Natural Science Foundation of China (NSFC, Nos.\ 12288102, 12333008, 12090040/1, 11873016, 11973080, 11803030), the National Key R\&D Program of China (Nos.\ 2021YFA1600403, 2021YFA1600401 and 2021YFA1600400), the Chinese Academy of Sciences (CAS), the Yunnan Ten Thousand Talents Plan–Young \& Elite Talents Project, and the CAS ‘Light of West China’ Program, the International Centre of Supernovae, Yunnan Key Laboratory (No. 202302AN360001), the Yunnan Fundamental Research Projects (grant Nos.\ 202401BC070007, 202201BC070003, 202001AW070007) and the ``Yunnan Revitalization Talent Support Program''—Science \& Technology Champion Project (No.~202305AB350003). 
\end{acknowledgements}

\appendix                  

\bibliographystyle{raa}
\bibliography{ref}
\label{lastpage}

\end{document}